\documentclass[]{spie}  

\usepackage{amsmath,amsfonts,amssymb}
\usepackage{graphicx}
\usepackage[colorlinks=true, allcolors=blue]{hyperref}
\usepackage{color}

\title{Enhanced high-dispersion coronagraphy with KPIC phase II: design, assembly and status of sub-modules}

\author{N.~Jovanovic$^{a}$, B. Calvin$^{a}$, M. Porter$^{a}$, T. Schofield$^{a}$, J. Wang$^{a}$, M. Roberts$^{a}$, G.~Ruane$^{b}$, J. K.~Wallace$^{b}$, R. Bartos$^{b}$, J. Pezzato$^{a}$, J. Colborn$^{a}$, J. R.~Delorme$^{a,c}$, D.~Echeverri$^{a}$, D.~Mawet$^{a,b}$, C. Z. Bond$^{d}$, S. Cetre$^{c}$, S. Lilley$^{c}$, S. Ragland$^{c}$,  P. Wizinowich$^{c}$, R. Jensen-Clem$^{d}$

$^{a}$ Department of Astronomy, California Institute of Technology, 1200 E. California Blvd., Pasadena, CA, 91125, USA; \\
$^{b}$ Jet Propulsion Laboratory, California Institute of Technology, 4800 Oak Grove Drive, Pasadena, CA, 91109, USA; \\
$^{c}$ W. M. Keck Observatory, 65-1120 Mamalahoa Hwy., Kamuela, HI, 96743, USA; \\
$^{d}$ Department of Astronomy \& Astrophysics, University of California, Santa Cruz, CA 95064, USA;\\
}

\authorinfo{Further author information: (Send correspondence to Nemanja Jovanovic)\\Nemanja Jovanovic: E-mail:  jovanovic.nem@gmail.com, Telephone: +1 626 395 1214}

\begin{document} 
\maketitle

\begin{abstract}
The Keck Planet Imager and Characterizer (KPIC) is a purpose-built instrument for high-dispersion coronagraphy in the K and L bands on Keck. This instrument will provide the first high resolution (R$>$30,000) spectra of known directly imaged exoplanets and low-mass brown dwarf companions visible in the northern hemisphere. 

KPIC is developed in phases. Phase I is currently at Keck in the early operations stage, and the phase II upgrade will deploy in late 2021. The goal of phase II is to maximize the throughput for planet light and minimize the stellar leakage, hence reducing the exposure time needed to acquire spectra with a given signal-to-noise ratio. To achieve this, KPIC phase II exploits several innovative technologies that have not been combined this way before. These include a 1000-element deformable mirror for wavefront correction and speckle control, a set of lossless beam shaping optics to maximize coupling into the fiber, a pupil apodizer to suppress unwanted starlight, a pupil plane vortex mask to enable the acquisition of spectra at and within the diffraction limit, and an atmospheric dispersion compensator. These modules, when combined with the active fiber injection unit present in phase I, will make for a highly efficient exoplanet characterization platform. 

In this paper, we will present the final design of the optics and opto-mechanics and highlight some innovative solutions we implemented to facilitate all the new capabilities. We will provide an overview of the assembly and laboratory testing of the sub-modules and some of the results. Finally, we will outline the deployment timeline. 

\end{abstract}

\keywords{Wavefront sensing, high contrast imaging, exoplanets, high dispersion coronography, high resolution spectroscopy}

\section{INTRODUCTION}
\label{sec:intro}  

The characterization of exoplanetary atmospheres demands innovative new approaches which improve on sensitivity and precision. High dispersion coronography (HDC) is one such approach that provides the ability to do species-by-species molecular characterization (e.g.~oxygen, water, carbon dioxide, methane), thermal (vertical) atmospheric structure, planetary spin measurements (length of day), and potentially Doppler imaging of atmospheric (clouds) and/or surface features (continents versus oceans)~\cite{wang2017OEH}. HDC optimally combines high contrast imaging techniques such as adaptive optics/wavefront control plus coronagraphy to high resolution spectroscopy~\cite{wang2017OEH,mawet2017OEH}. One approach that is being explored is to use an optical fiber to route the light of the known planet from the focal plane to the spectrograph. A single-mode fiber (SMF) is the ideal transport vehicle owing to the fact it has a field-of-view (FOV) that can be matched to the 1~$\lambda/D$ width of the point spread function (PSF), enabling efficient coupling for the planet light and suppression of unwanted star light. In addition, its narrow FOV reduces the amount of sky background injected into the spectrograph, which can be several orders of magnitude greater in the case of a seeing-limited spectrograph. 

Several projects have been initiated to realize this new technique from the ground including: the Keck Planet Imager and Characterizer (KPIC)~\cite{Mawet2016_KPIC,jovanovic2019-KPI}, which combines Keck AO and NIRSPEC, the Rigorous Exoplanetary Atmosphere Characterization with High dispersion coronography instrument (REACH)~\cite{jovanovic2017DPC}, which combines SCExAO and IRD and High-Resolution Imaging and Spectroscopy of Exoplanets (HiRISE), which combines SPHERE and CRIRES+\cite{vigan2018BHS}. The phase I version of KPIC and the REACH instrument are both transitioning from commissioning to early science at the time of writing of this article and offer complimentary wavelength coverage across the near-IR (NIR) on Maunakea (REACH operates from y-H and KPIC operates in K and L bands), while HiRISE is still in the development stage.

KPIC consists of a NIR pyramid WFS (PyWFS), which provides advanced wavefront correction and a Fiber Injection Unit (FIU), which injects the light into optical fibers, that are routed to NIRSPEC. KPIC is being developed in phases. Phase I consisted of the PyWFS and a basic FIU, which were deployed to Keck in the fall of 2018, have been commissioned and are now routinely used for scientific observations. This new NIR WFS provides exceptional correction, especially on redder targets too faint for the Shack-Hartmann~\cite{bond2020-AOW}. Phase II consists of a series of upgrades to the FIU, which aim to maximize the planet light coupled to the fiber, while minimizing the amount of stellar leakage (important for HDC) so as to minimize the integration time to reach a given SNR~\cite{pezzato2019-SKP}. Phase II is currently undergoing laboratory testing and will be deployed in the fall of 2021 to replace the phase I FIU optical layout at Keck. Here we provide an overview of the phase II design and present some of the laboratory characterization for the various sub-modules.

\section{Opto-mechanical design}

\begin{figure} [b!]
   \begin{center}
   \begin{tabular}{c} 
   \includegraphics[width=0.98\textwidth]{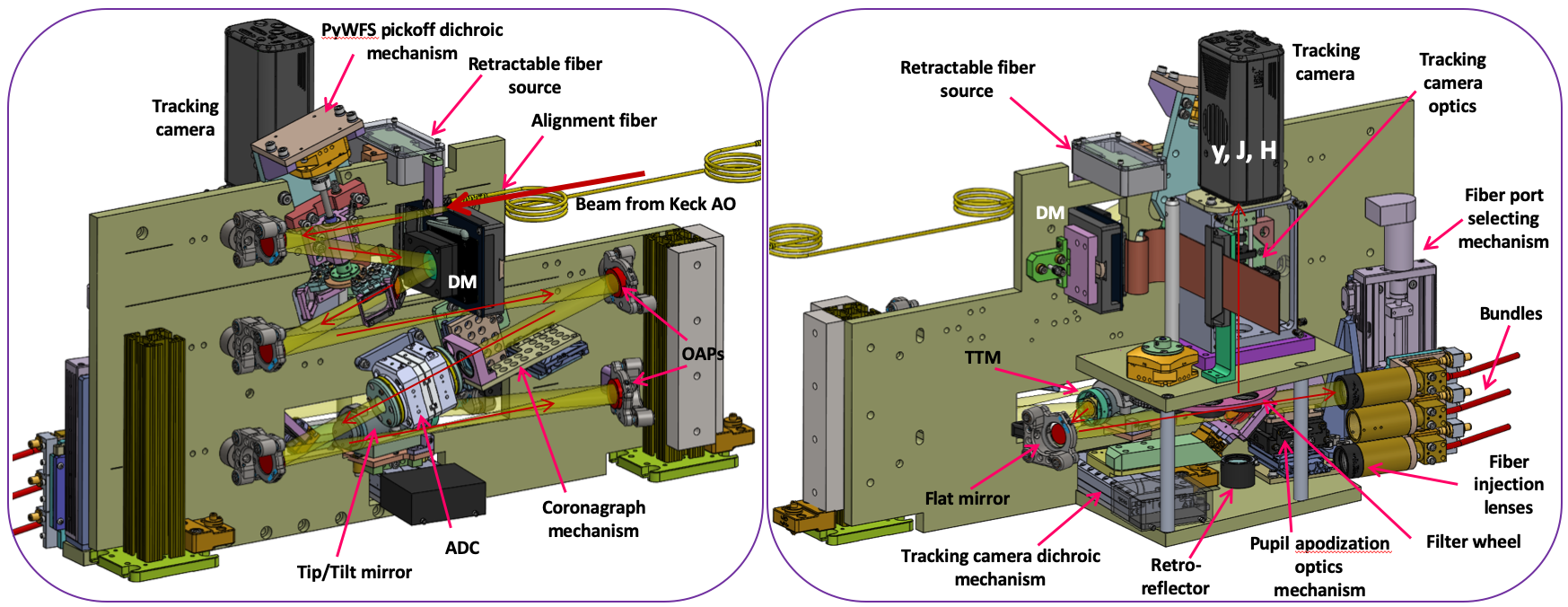}
   \end{tabular}
   \end{center}
   \caption[example] 
   { \label{fig:KPIC} 
A CAD drawing of the phase II version of the fiber injection unit for KPIC. (Left) View from the front. (Right) View from the rear. All major opto-mechanical modules are labelled, along with the direction of propagation of the beam through the system.}
\end{figure}

%
Owing to the fact that KPIC is located in a small space inside Keck AO, KPICs optics are mounted on a vertical plate. The plate is mounted on kinematics so that the phase I plate at Keck could be replaced with the phase II plate in late 2021. The opto-mechanical design of the phase II plate can be seen in Fig.~\ref{fig:KPIC}, which we will describe in detail. 

The converging beam on its way to NIRC2 is picked off and redirected to KPIC via three gold mirrors (not shown in~\ref{fig:KPIC}). The purpose of the KPIC plate is to provide a relay for the light to the final focal plane, which has an optimized F/\# set to maximize coupling to the fiber bundle, while providing sufficient focal/pupil planes for all the control and beam manipulation optics that maximize planet light coupling while minimizing star light leakage. To this end, the back bone of the KPIC design is a double optical relay that consists of two off-axis parabolic (OAP) mirrors. The OAPs are oriented the incorrect way with respect to the incoming beam. In this arrangement there are significant aberrations in the collimated beam spaces, but this helps preserve the field in the downstream focal planes. However, a single OAP relay used like this would induce a large field tilt and therefore a second relay is needed to flatten the field once again. A final OAP is then used (the correct way around) to recollimate the beam one last time before it is focused using a custom triplet lens (CaF$_{2}$, ZnS, AMTIR-1, Rainbow Research Optics) onto the fiber bundle.  

This optical arrangement creates three pupil planes. In the first pupil plane, a 1000 element deformable mirror (DM, Kilo-C-3.5, Boston Micromachines) can be seen. It has a protective window to prevent advanced aging, which can occur when operating the DM in a high humidity environment. The window is made from CaF$_{2}$ because it is transparent across the entire operating range of KPIC (y-L bands). The DM will provide a significant increase in the number of actuators compared to the native Keck AO DM, which has 349, increasing the control radius and reducing the fitting error in the correction. In addition, the DM will have lower latency electronics and can be driven at kHz rates also helping to offer a better adaptive optics correction. It will be mounted on a rail so that it can be easily removed/replaced when the plate needs to be transported.

The PyWFS pickoff is a watermill style wheel located immediately after the DM. The switching mechanism consists of a rotation stage (Conex-AG-PR100P, Newport) that drives the watermill like optics holder. To avoid mechanical collisions, the optics holder is offset from the rotator by a shaft and bearing housing. The mechanism is used to select the light utilized for wavefront control by the PyWFS. The reflected beam can be steered to adjust the pointing into the PyWFS. Each optic in the wheel can be adjusted in pitch by rotating the wheel and in yaw by a compact flexure-based adapter used to hold the optic cells to the hub of the wheel. This is useful for co-aligning the pointing of the various optics. However, to course align the entire mechanism, the sub-module is fitted with push/pull actuators and springs that enable the beam to be reflected accurately towards the PyWFS. 

As the PyWFS is sensitive from y-K band, it could in principle be used across this range, assuming the cold and warm filters in its beam path are upgraded from that of the phase I design. The pickoff mechanism provides 4 slots for optics. Three slots will be reserved for custom dichroics to share light with the PyWFS. One will be for an AR coated piece of glass that matches the physical dimensions of the other optics in the wheel. The purpose of this optic is to offset any alignment beam (pupil shift) as if it were passing through an actual beamsplitter, but allow for the alignment beam to be fully transmitted, unlike the dichroics. This optic is critical to alignment and maintenance of KPIC as identified in phase I.  

\begin{figure} [t!]
   \begin{center}
   \begin{tabular}{c} 
   \includegraphics[width=0.98\textwidth]{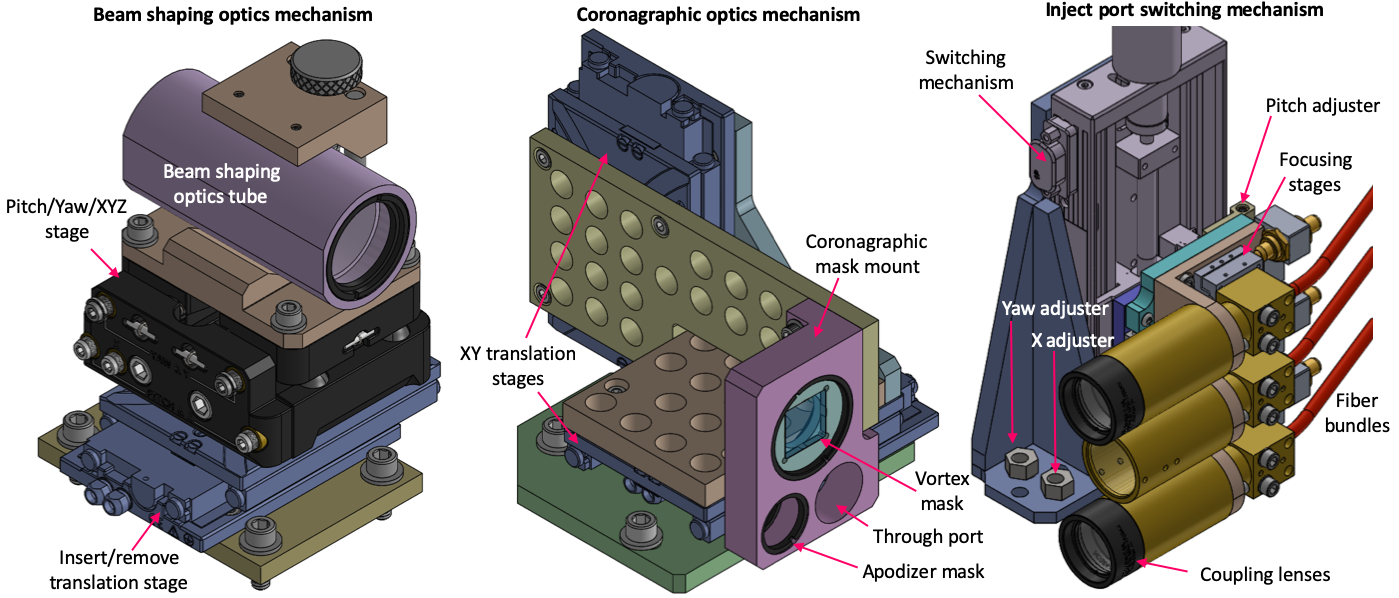}
   \end{tabular}
   \end{center}
   \caption 
   { \label{fig:PIAA} 
Several opto-mechanical assemblies. (Left) Beam shaping optics mechanism. (Center) Coronagraphic mask switching mechanism. (Right) Fiber injection port switching mechanism.}
\end{figure}

In the second pupil plane, a corongraphic module with a vector vortex mask~\cite{mawet2005-AGP} and a grey scale micro-dot apodizer (MDA)~\cite{zhang2018CMA} is located. See Fig.~\ref{fig:PIAA} for details. Both masks are used for suppressing starlight. The vortex mask is used to enable the vortex fiber nulling (VFN) mode of operation, which provides an inner working angle of $<1\lambda/D$~\cite{ruane2018-ESE,echeverri2019-TVF,ruane2019-VFN,echeverri2019-VFN}. Please see Ref.~\citenum{echeverri2020-PVC} of this conference for the latest update on this mode. At such close separations, KPIC could directly collect the spectra of many RV planets for the first time. On the other hand, the MDA is designed to suppress starlight in the focal plane from 2-10$\lambda/D$ at the expense of throughput. The coronagraphic mechanism consists of two linear stages (Q-545.241, PI), which allows the coronagraphic masks to be inserted into and aligned with the beam. For alignment purposes, the tracking camera (outlined below) can be used in pupil-viewing mode. Both masks operate without Lyot stops downstream.   

Immediately after the coronagraphic masks is an atmospheric dispersion corrector (ADC), which aims at correcting for the differential atmospheric refraction and restoring the PSF, to a single spot, essential to boosting coupling to the fibers. The module consists of 2 sets of 3 prisms (BaF$_{2}$, CaF$_{2}$, ZnSe), which counter rotate with respect to the elevation axis of the telescope. Two compact rotation stages (PR-50, Micronex) are used to allow the prisms to rotate with respect to one another and track the elevation axis, which rotates in the pupil in fixed field mode. Each stage uses a magnetic encoder to keep track of its position. A lookup table of rotation angle offsets with respect to the elevation axis for each prism as a function of Zenith angle is used to drive the stages during observations. This table is generated through detailed simulations of the properties of the atmosphere and prisms. The prisms in the mechanism are optimized for operation across J to L band. More details about the optical design and performance of the ADC can be found in Ref.~\citenum{wang2020-AAD}.

\begin{figure} [t!]
   \begin{center}
   \begin{tabular}{c} 
   \includegraphics[width=0.98\textwidth]{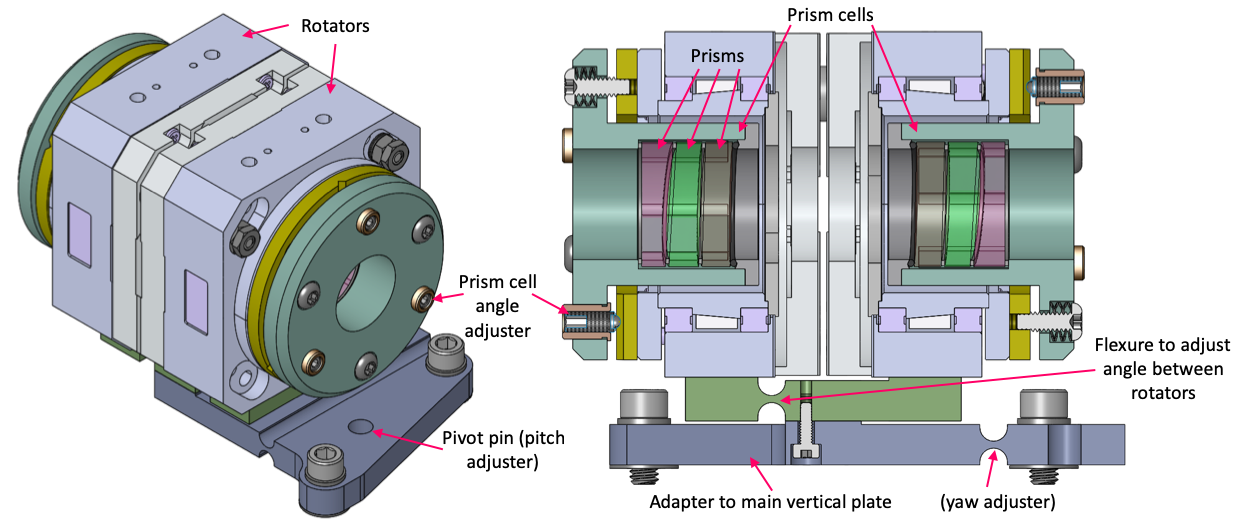}
   \end{tabular}
   \end{center}
   \caption[example] 
   { \label{fig:ADC} 
The ADC opto-mechanical assembly. (Right) A cross-sectional view showing the prisms mounted in their cells, the cells mounted in the rotation stages, and the stages mounted on the adapter plates. The angle of the prisms with respect to the axis of rotation can be adjusted with the plungers. The angle between the two stages can be adjusted with the flexure plate at the bottom and the spacers at the top, and the angle of the entire assembly with respect to the beam on the master plate can be adjusted using the pivot pin and the flexure based adapter.}
\end{figure}

Figure~\ref{fig:ADC} shows some CAD images of the ADC mechanism. The prisms are mounted in custom machined cells. Their separation and parallelism is maintained with precision shims, while their clocking is based on the prisms each having a flat side (i.e. D-shaped) pressed against a reference flat in the cell. The cell is secured to the stage with spring loaded fasteners which allow for slight adjustments in the pitch and yaw so the input facet of each prism set can be aligned perpendicular to the optical axis. Once this step is complete, the rotators need to be secured to one another so the rotation axis of the two motors is parallel. This can be achieved by using a flexure-based mounting plate as shown in the figure. By adjusting the thickness of the washers for the two threaded bolts used to keep the opposite end of the rotators together, its possible to adjust the co-alignment of the two mechanisms. Once assembled, the entire assembly will be located with a pin on the master plate. The far end of the mechanism can be swiveled about this pin to adjust pitch and the thickness of the spacers/washers under the far end can be used to adjust yaw. In this way, all the degrees of freedom needed to align the ADC mechanism can be provided in an extremely compact form factor.

A tip/tilt mirror (S-330.8SL, PI) is located in the final pupil plane of the system. The TTM is used to steer the beam in the focal plane to align it with the core of the optical fibers in the bundle. It has two adjusters that allow the reflected beam to be coarsely aligned in the right direction. The purpose of the adjusters is to allow for the bulk TTM assembly to be steered to align the beam with the tracking camera and back end optics without using precious range on the TTM itself. To do this, the TTM will be set to the middle of its range in X and Y and then be steered using the course alignment mechanisms. A gold coated mirror, is attached to the TTM mechanism via a flexure-based holder. The holder applies the lightest pre-load on the mirror laterally, preventing deformations and hence unwanted aberrations.   
 
The beam is next incident on a flat mirror, which directs it towards the tracking camera pickoff (TCP) and ultimately the fiber injection. The TCP consists of a linear translation stage (N-565.360, PI) that can switch between three optics. One is used for alignment as was the case for the PyWFS pickoff. The other two can be freely chosen to split the light between tracking and science as needed. The TCP optics are mounted in a single mount, which can be aligned in pitch and yaw. However, there is no way to adjust the differential alignment between the beams reflected by the three optics. This is acceptable because the optics are co-mounted so the differential pointing is very small (a few $\lambda/D$) and the FOV of the tracking camera is relatively large ($\sim165\times132\lambda/D$).  

\begin{figure} [t!]
   \begin{center}
   \begin{tabular}{c} 
   \includegraphics[width=0.98\textwidth]{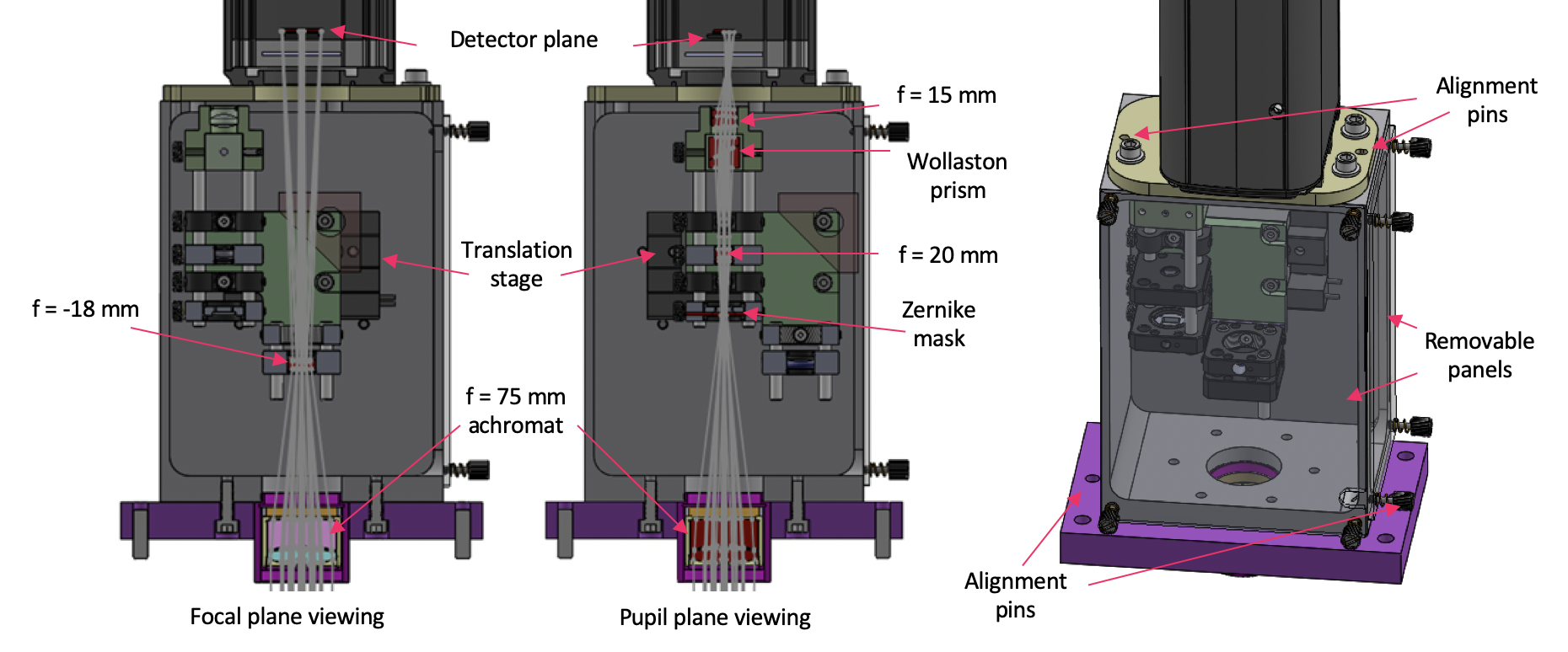}
   \end{tabular}
   \end{center}
   \caption[example] 
   { \label{fig:trackcamera} 
The tracking camera opto-mechanical assembly. (Left) Focal plane viewing configuration. (Center) Pupil plane viewing configuration. (Right) The camera is mounted on a plate, which is pinned to the optics block. The entire block is pinned to the master plates for easy removal and replacement for servicing.  }
\end{figure}

The light reflected by the dichroics is directed to the tracking camera (C-red2, FirstLight Imaging), which has several observing modes. These include a focal plane viewing mode, used for tracking targets and NIR imaging, as well as a pupil plane viewing mode and a Zernike WFS mode. The internal optics inside the tracking camera module can be seen in Fig.~\ref{fig:trackcamera}. The focal plane viewing mode provides a beam with an F/\#=39, which results in a plate scale of $\sim8$ mas/pixel and hence almost 3 pixels/FWHM at the short end of J-band. The focal plane viewing mode is used to track the star and guide on the fiber, while the beam transmitted by the TCP dichroic goes directly to a focusing optic that is used to inject the light into the fiber bundle. The tracking camera images are used to drive the TTM in KPIC to steer the beam to the fiber and maintain alignment. A corner cube is used to steer light reverse injected from the bundles onto the tracking camera so that the position of the bundle can be determined with respect to the beam from Keck AO.

A linear stage is used to replace the optics in the beam path and switch between the focal and pupil plane imaging configurations. In the pupil plane viewing mode, the light passes through a Wollaston prism, which generates two images of the pupil with orthogonal linear states of polarization. This is necessary for the Zernike WFS mode and to avoid having two different modes, the classical pupil plane imaging mode shares these optics. Each pupil plane image subtends $\sim200$ pixels on the C-red2. Given the Keck pupil has seven segments across the primary, this provides about $28$ pixels/segment, which is grossly over sampled, but is convenient for fully understanding the Zernike WFS as well as providing a precise reference image for aligning the coronagraphic masks to. The Zernike WFS is based on a mask with a phase dimple placed in the intermittent focal plane of the tracking camera shown in Fig.~\ref{fig:trackcamera}. The phase dimple is simply a pit in a glass substrate that injects a $\pi$ phase shift for the light within the FWHM of the PSF with respect to the light outside this. This helps convert phase aberrations into intensity modulations the C-red2 can see. KPICs Zernike WFS is a vector mask, which means it does this for both polarizations, which should provide a superior sensitivity. To activate this mode, the TTM is used to steer the beam onto the phase dimple, while classical pupil imaging is conducted by passing the beam through the optic away from the dimple. The Zernike WFS is an experimental module that could be used to help improve the co-phasing for the primary mirror of Keck, which would be an important demonstration for TMT and the ELT as well. 

To improve access for servicing the camera, the C-red2 is mounted on a plate, which is pinned to the optics block. To allow much easier mask replacements in future, the entire assembly is pinned to the master plates. It can be easily removed so masks can be exchanged outside of Keck AO and replaced without needing a complex realignment.  

In the beam path to the tracking camera is a filter wheel. This has 8 slots which includes filters such as a J, H, and 1550 nm bandpass filter that has a 25 nm bandwidth. There will also be an open slot, and a block so dark's can be taken. 

The beam transmitted by the tracking camera pickoff is next incident on the beam shaping, or PIAA optics~\cite{guyon2003-PIA}. These lenses reshape the pupil illumination from a flat-top to a quasi-Gaussian, which improves coupling into the SMF~\cite{jovanovic2017-EIL}. An aspheric lens pair is needed to achieve this, which is mounted into a fixed tube. The tube is mounted on a translation stage (Q-545.241, PI) used to move it in/out of the beam as well as a 5-axis manual stage, that can be used to carefully align the tube initially with the beam. In this way, the PIAA is deployable and retractable. The mechanism can be seen in Fig.~\ref{fig:PIAA}. The PIAA lenses in KPIC phase II are optimized for operation in the K and L bands, the primary science bands for HDC with KPIC. 

The final module is the injection port switching mechanism, which can be seen in Fig.~\ref{fig:PIAA}. Lenses are used to focus the beams onto the bundles with the optimum F/\#. KPIC Phase II will provide 3 ports. The ports mechanically align the optical axis of the lenses with the bundles. The bundles can only be translated in focus with respect to the lenses to maximize coupling (i.e. position them at the focus of the lenses). Manual actuators which push compact linear stages are used to achieve this and once aligned are fixed. A large translation stage (VT-80, PI) is used to switch the port that is inserted into the beam. The entire assembly can be adjusted manually in X/Y, pitch/yaw, to align the ports to the optical axis of the instrument. This gets close, but the final alignment of pitch/yaw is done with the aid of the TTM in KPIC.  

KPIC phase II will offer a retractable light source, which can be deployed at the input focal plane. This source will allow us to inject calibration light at the input to KPIC for daytime testing and calibration purposes. This means KPIC can receive light and can undergo testing while others are using Keck AO, increasing the amount of time KPIC can be used for laboratory testing. This was an upgrade identified by operating the phase I plate.   

As a final note, all opto-mechanical mechanisms were designed with the following constraints in mind: precision, accuracy, translation range, footprint, speed, heat dissipation of $<$10~W each, operational temperatures down to 0C on average, larger controllers could be placed 10 m away in the electronics room, control was compliant with a Linux, Centos07 OS, no visible encoder lights that could interfere with the WFS's in Keck AO and if possible the mechanisms should be a COTS parts. This greatly limited the number of possible options for each application. The final devices reported here met these requirements.

\section{Assembly and status}
\begin{figure} [b!]
   \begin{center}
   \begin{tabular}{c} 
   \includegraphics[width=0.98\textwidth]{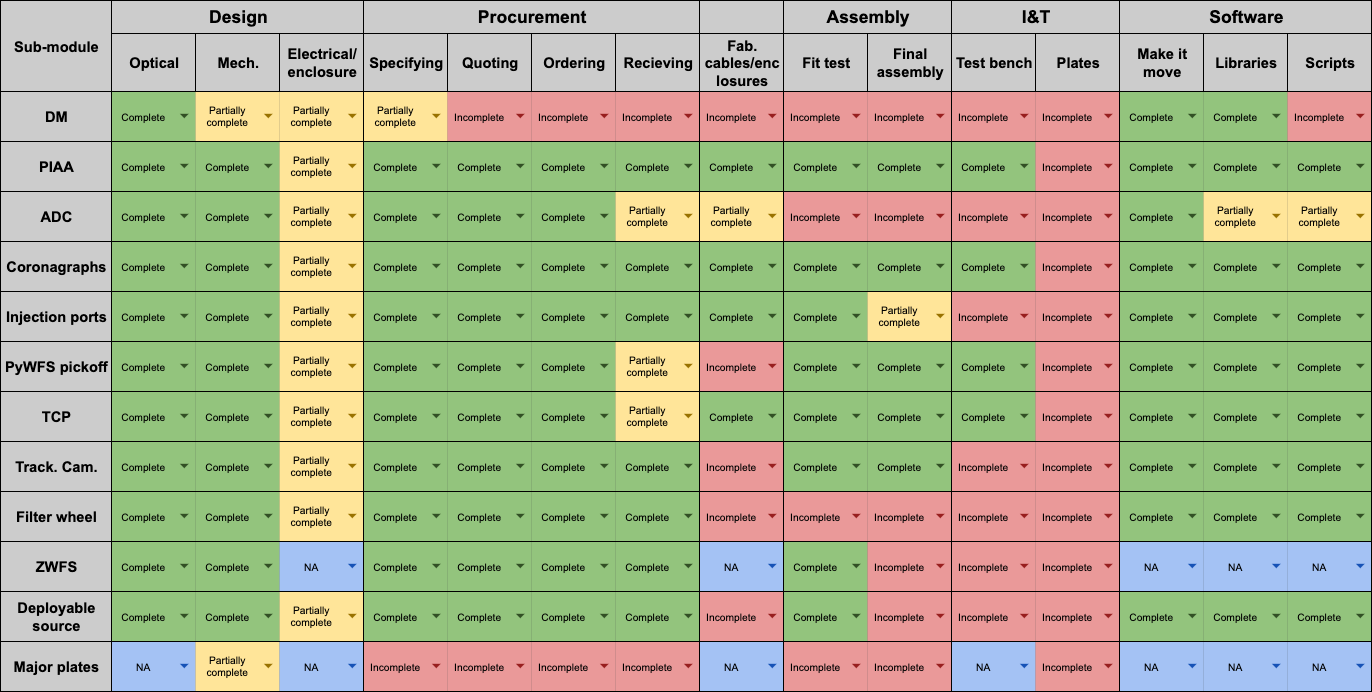}
   \end{tabular}
   \end{center}
   \caption 
   { \label{fig:status} 
Status of the development of the phase II sub-modules. Green means complete, red incomplete, yellow partially completed and blue not applicable. ZWFS refers to the Zernike WFS. I\&T refers to integration and testing. Test bench refers to custom optical bench where all modules are tested in isolation initially. Plates means that the module has been integrated onto the master plate as it will be at Keck and is undergoing testing at the system level. Make it move refers to initial attempts to communicate with opto-mech. by any means possible.}
\end{figure}

The status of the phase II sub-modules at the time of writing of this article is summarized in Fig.~\ref{fig:status}. Each sub-module as well as the master plates everything is bolted to have a separate row in the table. The development process flows through design, procurement/fabrication, assembly, testing and integration and control software.  
It is clear to see from the figure that most sub-modules have completed optical design, mechanical design, procurement, fit testing and the final assemblies have been built. In addition, control software has advanced beyond the readiness of the mechanics and all mechanisms have been made to move, and have low level libraries and higher level scripts to control sequenced motions that can operate in Linux Centos07, the OS of choice for Keck computer systems. Having access to control software at such an advanced stage is beneficial to efficient I\&T.

Once assembled, the sub-modules will undergo testing in two stages. Stage 1) involves integrating the devices into a custom optical setup so they can be tested in isolation initially, referred to as ``Test bench" in the figure. Once the basic parameters of the module have been tested in this setup, it will be moved to the final assembly and integrated in the plates. This will allow system level testing needed to fully qualify the module before deployment. Once fully assembled, control scripts that aid acquisition, as well as various forms of wavefront control will be produced and tested in the laboratory.

\begin{figure} [t!]
   \begin{center}
   \begin{tabular}{c} 
   \includegraphics[width=0.98\textwidth]{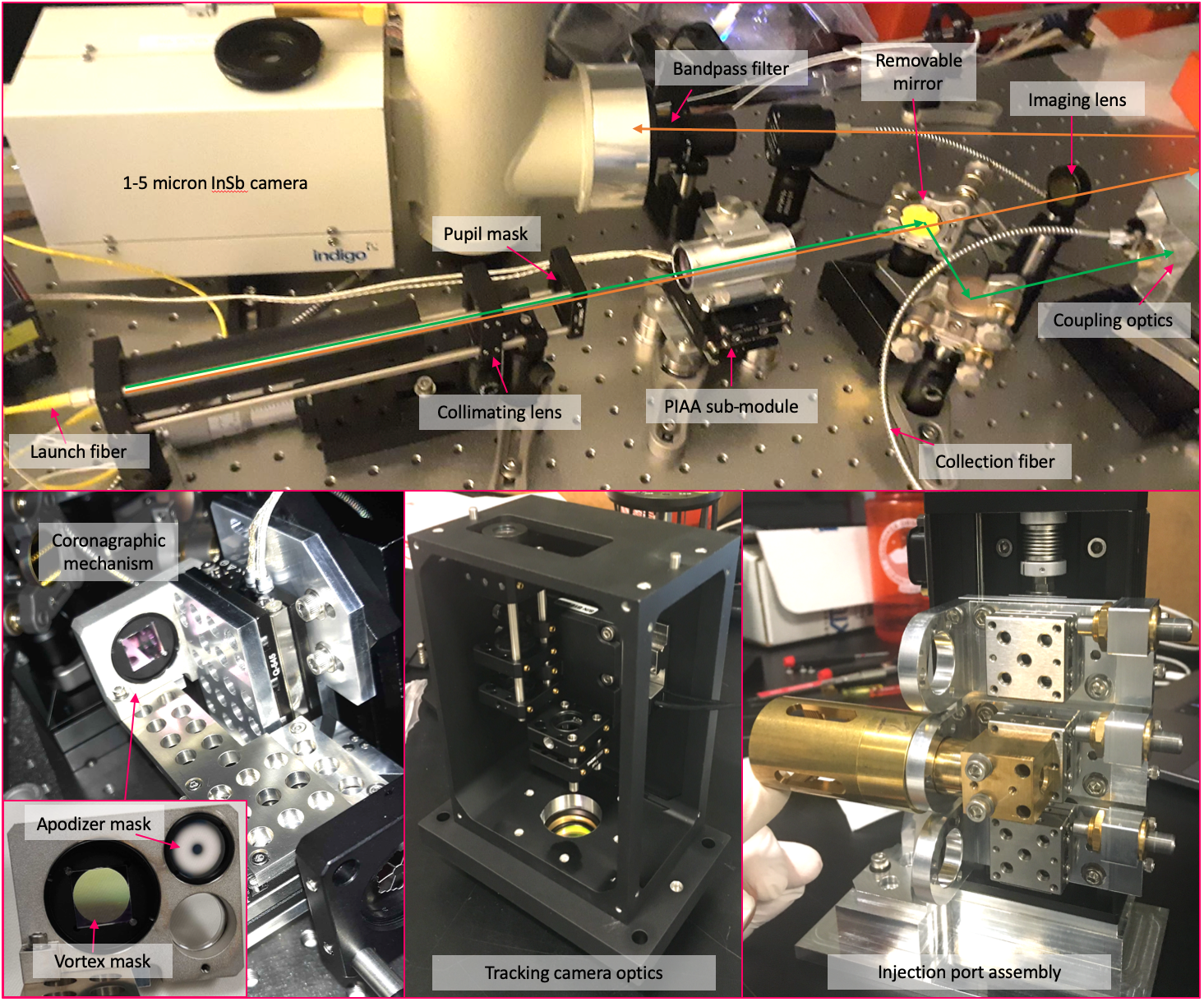}
   \end{tabular}
   \end{center}
   \caption 
   { \label{fig:modules} 
Images of assembled sub-modules. (Top) The test bench used to test each sub-module in isolation. The beam shaping (PIAA) sub-module is shown in the center of the bench undergoing testing. The bench consists of a beam from an optical fiber which is collimated and then passed through a mask that replicates the Keck pupil, passes the beam through the optic under test before either directing it towards a NIR camera used to image the beam (orange path), or towards coupling optics used to inject the light into a SMF (green path). These can be switched by removing/replacing a mirror on a magnetic mount. (Lower left) The coronagraphic mechanism and the masks mounted in their holder. (Center) The tracking camera optics block. (Right) The injection port switching mechanism with one optic mounted.}
\end{figure}

Images of some of the fully assembled modules are shown in Fig.~\ref{fig:modules}. It can be seen that the modules that have been tested on the test bench include the beam shaping optics, the coronagraphs, the PyWFS pickoff and TCP. These were accelerated because they posed a higher level of risk and needed to be qualified early on to prevent delays down the track. The ADC is the only other high risk element that has not been assembled or tested yet, owing to the difficulty in locking down vendors to produce and AR coat the prisms. The optics are about to be delivered and assembly and testing will begin in the near future. 

The timeline for the remainder of the development is to have all modules fully assembled and tested in the test bench by March 2021. They will then be integrated into the plates and system level testing will begin. The project will have a pre-ship review in late September 2021 and if successful, the phase II plate will be deployed to the Keck II telescope in November.

\section{Preliminary laboratory characterization}~\label{sec:results}
As seen in Fig.~\ref{fig:status}, four sub-modules have been characterized in the laboratory, including the beam shaping optics, the coronagraphs, the PyWFS pickoff and the TCP. The optical performance, including the shape of the PSF, the throughput of the optic and the coupling to a SMF for the coronagraphs and PIAA optics is presented in~\citenum{calvin2020-EDS} and will not be repeated here. Here we instead report on some of the key results for the other two mechanisms instead.   

\subsection{The PyWFS pickoff mechanism}
The PyWFS pickoff is used to select the dichroic that will split the light between the PyWFS and the FIU. A critical property of the mechanism is how accurately the beam reflected towards the PyWFS could be aligned to the optical axis, and its stability. To test this, mirrors were placed in each of the four optic cells. A collimated laser beam was reflected from one of the mirrors and imaged onto a detector. The centroid of the beam was noted and the wheel rotated to the next mirror. At this position, the beam was co-aligned with the position of the first by using the wheel and the flexure mechanism mentioned above. This process was repeated for all four mirrors. This demonstrated that there was sufficient range in the flexure mechanism to be able to accurately co-align all four beams, arresting that concern. 

\begin{figure} [h!]
   \begin{center}
   \begin{tabular}{c} 
   \includegraphics[width=0.99\textwidth]{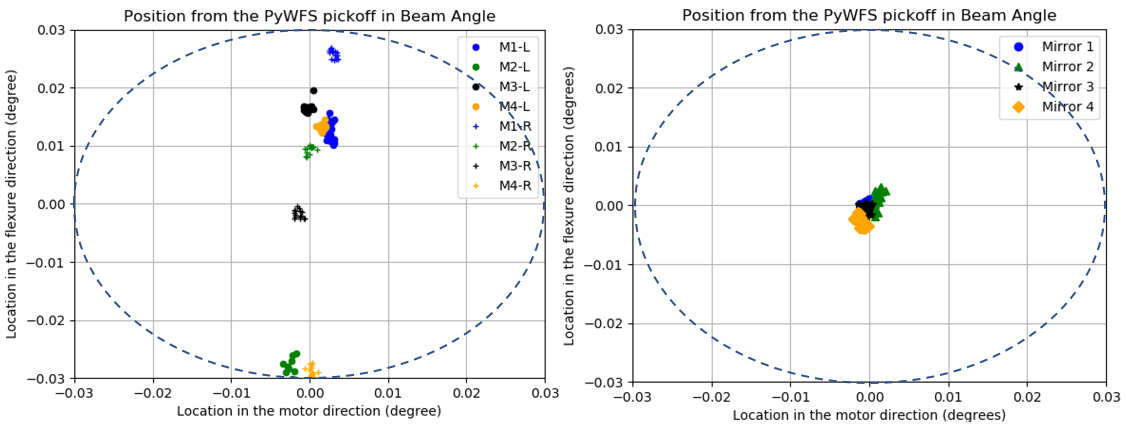}
   \end{tabular}
   \end{center}
   \caption 
   { \label{fig:pywfsp} 
Scatter plot of the position of the centroid each time a mirror is inserted into the beam with the PyWFS pickoff mechanism. (Left) Without and (Right) With compensating for the hysteresis effect.}
\end{figure}

Next, a script was executed that rotated the wheel from position 1 to 2, 2, to 3, 3 to 4, 4 to 3, 3 to 2, 2 to 1 on repeat to see how the centroids would behave. The results are shown in the left panel of Fig.~\ref{fig:pywfsp}. It can be seen that for each mirror, there are two unique clusters of centroids depending on whether you approach that mirror from the left or the right (i.e. 1 to 2, or 3 to 2). After much testing it was determined that there was some hysteresis-like effect and that if one were to approach mirrors 3 and 4 directly from the home position, and mirrors 1 and 2 from the maximum angle of the stage (340$^{\circ}$), then the scatter could be minimized as is shown in the right panel of Fig.~\ref{fig:pywfsp}. This shows a significant reduction in the spread of the spots, which results in a much more deterministic and reproducible alignment of the PyWFS beam as one switches between optics. For reference the blue dashed circle shows the requirement for the maximum misalignment with respect to the optical axis allowed and the figure clearly shows that even without hysteresis compensation, the mechanism meets specification.     

\subsection{The tracking camera pickoff mechanism}
\begin{figure} [h!]
   \begin{center}
   \begin{tabular}{c} 
   \includegraphics[width=0.98\textwidth]{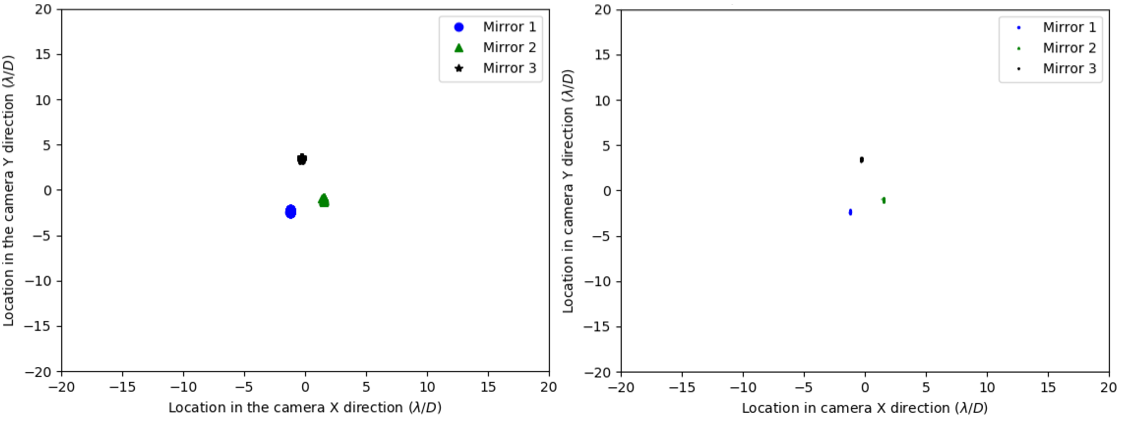}
   \end{tabular}
   \end{center}
   \caption 
   { \label{fig:TCP} 
Scatter plot of the position of the centroid each time a mirror is inserted into the beam with the TCP mechanism. (Left) Centroids and (Right) RMS spread of centroids for each mirror.}
\end{figure}

The TCP will be used to split the light between the tracking camera and the fiber injection ports. Although the tracking camera has a large FOV (5.5$\times$4 arcsec), its important to make sure that the centroids from the three co-mounted optics are aligned to within the FOV of the camera and their pointings are highly repeatable. In a similar fashion to the PyWFS pickoff tests, a laser beam was bounced off a mirror in the TCP and an image formed. The centroid was recorded, the TCP mechanism switched to another mirror position and the centroid recorded again. This process was repeated 240 times over a period of 33 hrs. The results are shown in Fig.~\ref{fig:TCP}. It can be seen that each mirror points in a unique direction, but that the spread in pointing after successive moves away from and back to a given mirror is very small (see right panel). This highlights that the pointing is highly repeatable. The three mirrors have an offset in their pointing of the order of $\sim5\lambda/D$, or 40 mas or 5 pixels on the C-red2. Given the size of the array (640$\times$512), this is a minor offset and well within the central region of the camera. As such, the mechanism can successfully co-point the three beams and provides highly reproducible pointing to warrant application in KPIC phase II.  

\section{Summary}~\label{sec:summary}
The KPIC instrument is already offering enhanced exoplanet characterization capabilities at Keck. Phase II will provide a series of sub-module upgrades that will improve planet throughput and reduce stellar leakage, ultimately reducing overall integration times. In this paper we have provided a detailed overview of the entire opto-mechanical design and its rationale. The phase II modules are partly built and testing is on-going in the laboratory. The goal is to complete all sub-modules in early 2021, and move to integrating them into the system. This will provide a 6 month window to verify, validate and develop control scripts so the upgrade can be deployed in late 2021.

\acknowledgments 
 
This work was supported by the Heising-Simons Foundation through grants \#2019-1312 and \#2015-129. G. Ruane was supported by an NSF Astronomy and Astrophysics Postdoctoral Fellowship under award AST-1602444. We thank Dr. Rebecca Jensen-Clem for loaning AOSE for use within the KPIC phase II testing. Part of this work was carried out at the Jet Propulsion Laboratory, California Institute of Technology, under contract with the National Aeronautics and Space Administration (NASA). W. M. Keck Observatory is operated as a scientific partnership among the California Institute of Technology, the University of California, and the National Aeronautics and Space Administration (NASA). The Observatory was made possible by the generous financial support of the W. M. Keck Foundation.The authors wish to recognize and acknowledge the very significant cultural role and reverence that the summit of Maunakea has always had within the indigenous Hawaiian community. We are most fortunate to have the opportunity to conduct observations from this mountain.

\bibliography{report} 
\bibliographystyle{spiebib} 

\end{document}